\documentclass{article}

\usepackage[preprint]{neurips_2026}

\usepackage[utf8]{inputenc}
\usepackage[T1]{fontenc}
\usepackage{hyperref}
\usepackage{url}
\usepackage{booktabs}
\usepackage{longtable}
\usepackage{array}
\usepackage{calc}
\usepackage{amsmath}
\usepackage{float}
\usepackage{graphicx}
\usepackage{placeins}
\usepackage{xcolor}

\providecommand{\tightlist}{%
  \setlength{\itemsep}{0pt}\setlength{\parskip}{0pt}}
\providecommand{\pandocbounded}[1]{\resizebox{\linewidth}{!}{#1}}

\title{Peer-Voted LLM-Agent Stress Tests Find Feed-Induced Lexical
Convergence but No Reliable Matched-Exposure Advantage for Distributed
Sources}
\author{Rana Muhammad Usman
  \ \texttt{usmanashrafrana@gmail.com}
  \and
  Dominic Williamson
  \ \texttt{dominic@codarossa.ai}}

\begin{document}

\maketitle

\paragraph{Code and data availability.}
The complete implementation, frozen protocol, configurations, summary
tables, and paper source are available at
\url{https://github.com/ranausmanai/synthetic-social-networks}. Raw traces
are available at
\url{https://huggingface.co/datasets/ranausmans/synthetic-social-networks}.
The release contains 59,776 production posts from 528 trials: the 448-trial
confirmation and 80-trial exploratory stage. It includes every confirmatory
post, peer vote, exposure record, and agent endpoint.

\begin{abstract}
Population-level behavior in large-language-model (LLM) agents cannot be
characterized by single-agent benchmarks. We introduce PV-SST, a
peer-voted social-platform testbed, and report a separately frozen,
preregistered matched-exposure experiment spanning four topics, four
unused seeds, four open-weight model families, and three prespecified
larger variants. The experiment comprises 448 trials and 112 complete
model-by-topic-by-seed blocks. Relative to a topic-only control, a feed
of previous-round peer posts ranked by peer-generated likes increases
final-round lexical similarity in both the four-family core panel
(paired mean difference +0.0082 TF-IDF cosine units, 95\%
block-bootstrap CI {[}0.0043, 0.0121{]}, randomization p=0.000105, n=64
blocks) and the three-variant size extension (+0.0109 {[}0.0069,
0.0151{]}, p=0.000001, n=48). This contrast bundles peer-post exposure
with ranking and therefore does not identify a ranking-only effect.
Opposite-side survival falls in the core panel (-3.9 percentage points
{[}-6.8, -1.6{]}, p=0.0068) but not conclusively in the larger variants
(-1.0 pp {[}-3.1, 0.4{]}, p=0.50). Holding adversarial impressions
fixed, four distributed sources do not reliably move honest-agent stance
more than one source. The preregistered distributed-minus-single
contrast is positive but inconclusive in the core panel (+0.057
{[}-0.009, 0.125{]}, p=0.112) and negative in the larger variants
(-0.040 {[}-0.113, 0.035{]}, p=0.332), failing the prespecified
cross-model and cross-topic consistency criterion. Thus the robust
result is lexical convergence under the tested peer-ranked feed, not
general opinion capture or a general coordination advantage. The study
evaluates synthetic LLM-agent populations; it does not estimate effects
on people or production platforms.
\end{abstract}

\section{Introduction}\label{introduction}

LLM agents increasingly share environments as assistants, simulated
users, customer-service systems, and autonomous participants. Existing
evaluations mostly measure isolated responses or task completion.
Platform behavior, however, is recursive: agents produce content, other
agents react, platforms rank the resulting signals, and those signals
shape later generations. Single-agent benchmarks do not measure this
loop.

This gap matters for two common platform-risk claims. First, engagement
surfaces are often treated as passive wrappers around content, even
though ranked exposure may change the population before an attacker
appears. Second, coordinated campaigns are often described as
intrinsically more persuasive than individual sources. In observational
settings, coordination is entangled with reach, message volume,
targeting, and network position. Without matched exposure, account count
cannot be separated from these other advantages.

Human-subject experiments establish that popularity signals can
influence behavior \citep{salganik2006,muchnik2013}, and classical
opinion dynamics models formalize social influence
\citep{degroot1974,friedkin1990,hegselmann2002}. Neither directly
answers how language-producing LLM-agent populations behave under
endogenous peer feedback. Recent LLM social simulations establish the
feasibility of persona-conditioned populations and large-scale diffusion
\citep{park2023,chuang2024,yang2024}, but mechanism isolation and
run-level auditability remain open evaluation problems.

We address these problems with the \textbf{Peer-Voted Social Simulation
Testbed (PV-SST)}. Persona-conditioned agents generate short posts, cast
structured in-character votes, and receive platform-mediated feedback in
later rounds. Every post, vote, exposure, stance update, and parse
status is preserved. We first used a small exploratory study to identify
candidate mechanisms. We then froze a new protocol before inspecting
confirmatory outcomes and ran a 448-trial matrix using seven model
variants across four families.

The paper makes three contributions:

\begin{enumerate}
\def\labelenumi{\arabic{enumi}.}
\tightlist
\item
  \textbf{An auditable population-evaluation instrument.} PV-SST makes
  social feedback endogenous to the LLM-agent population while retaining
  complete run-level traces and deterministic request seeds.
\item
  \textbf{A matched-exposure coordination test.} One-source and
  four-source attacks deliver exactly the same number of adversarial
  impressions to the same honest population. The tested
  distributed-source package has no reliable general advantage.
\item
  \textbf{A multi-family confirmatory result.} A peer-ranked feed
  increases final-post lexical similarity across both model-size panels
  and every tested topic. Minority-view suppression, by contrast, is
  model-dependent and does not replicate conclusively in larger
  variants.
\end{enumerate}

These are statements about the implemented LLM-agent system, not human
behavior. The scientific role of the testbed is to falsify or qualify
threat models under controlled conditions before costly or risky human
studies.

\section{Related Work}\label{related-work}

\textbf{LLM-agent simulations.} Generative Agents demonstrated
persistent persona-conditioned social behavior \citep{park2023}.
Subsequent systems study opinion formation, diffusion, polarization, and
platform-scale interaction \citep{chuang2024,yang2024,zhang2025}. MOSAIC
combines social graphs, engagement actions, and misinformation
interventions \citep{liu2025}, while recent work examines conformity and
spiral-of-silence behavior in LLM collectives \citep{zhong2025}. PV-SST
does not claim to be the first social simulator. Its contribution is a
compact, peer-voted, fully logged testbed organized around paired
mechanism tests.

\textbf{Opinion dynamics and social proof.} DeGroot, Friedkin-Johnsen,
and bounded-confidence models provide explicit update rules for
convergence and influence
\citep{degroot1974,friedkin1990,hegselmann2002}. Human experiments show
that popularity signals can alter evaluation and consumption
\citep{salganik2006,muchnik2013}. These traditions motivate the
treatments but do not provide effect-size priors for LLM-agent
populations.

\textbf{Evaluation scope.} Large simulations and interview-conditioned
agents raise distinct questions of scale and behavioral fidelity
\citep{park2024,yang2024}. Our aim is narrower: internal causal contrast
within a synthetic system. We therefore treat model-topic-seed runs as
the inferential units, report model and topic strata, and avoid treating
thousands of generated posts as independent samples.

\section{PV-SST Confirmatory Design}\label{pv-sst-confirmatory-design}

\begin{figure}[t]
\centering
\pandocbounded{\includegraphics[keepaspectratio,alt={PV-SST protocol and paired contrasts}]{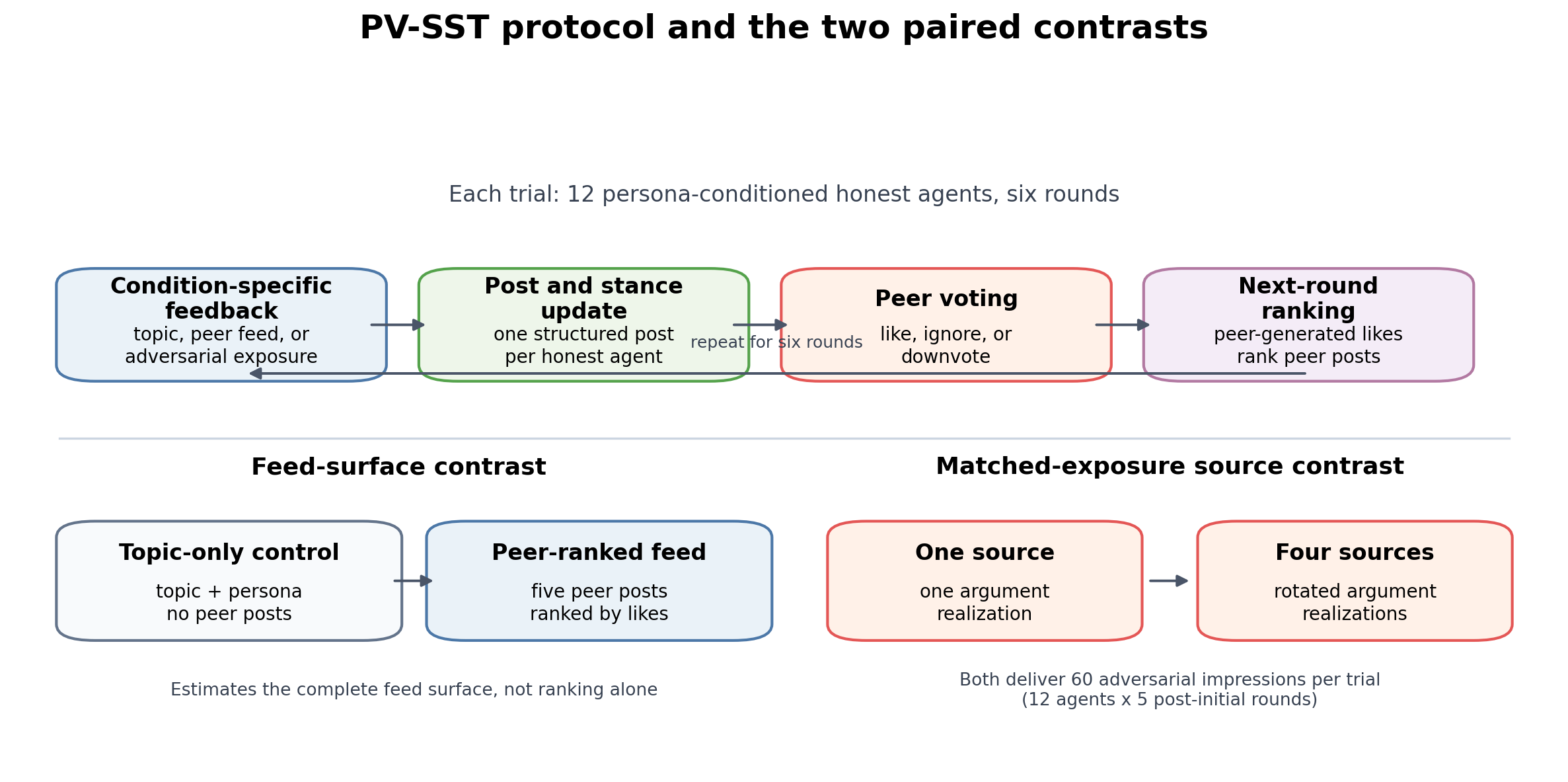}}
\caption{PV-SST protocol and paired contrasts. Each trial runs 12 persona-conditioned honest agents for six rounds. The feed contrast compares a topic-only control with five previous-round peer posts ranked by peer-generated likes, so it estimates the complete feed surface rather than ranking alone. The source contrast holds adversarial exposure fixed at 60 impressions per trial (12 agents x 5 post-initial rounds) while comparing one source with four distributed sources.}
\end{figure}
\FloatBarrier

\subsection{Testbed loop}\label{testbed-loop}

Each trial contains 12 honest agents sampled from a 24-persona pool.
Initial stances are exactly balanced across six non-neutral levels from
strongly support to strongly oppose, with common initial confidence
0.70. Each trial runs for six rounds. In every round:

\begin{enumerate}
\def\labelenumi{\arabic{enumi}.}
\tightlist
\item
  Each agent receives a condition-specific feedback block.
\item
  The agent emits a JSON-structured post of at most 180 characters and
  updates its discrete stance and confidence.
\item
  Each honest agent votes on up to five other posts using \texttt{like},
  \texttt{ignore}, or \texttt{downvote}.
\item
  The resulting peer votes determine the ranking shown in the next
  round.
\end{enumerate}

Generation and voting use Ollama with reasoning-token emission disabled.
Request seeds are deterministic functions of trial seed, round, phase,
and agent ordinal. Confirmatory votes preserve voter, target, integer
vote, and parse status; the compact frozen protocol omits free-text vote
rationales.

\subsection{Factorial matrix}\label{factorial-matrix}

The confirmatory protocol was frozen in
\texttt{CONFIRMATORY\_PREREGISTRATION.md} before outcomes were
inspected.

\begin{longtable}[]{@{}
  >{\raggedright\arraybackslash}p{(\linewidth - 2\tabcolsep) * \real{0.5000}}
  >{\raggedright\arraybackslash}p{(\linewidth - 2\tabcolsep) * \real{0.5000}}@{}}
\toprule\noalign{}
\begin{minipage}[b]{\linewidth}\raggedright
Factor
\end{minipage} & \begin{minipage}[b]{\linewidth}\raggedright
Confirmatory levels
\end{minipage} \\
\midrule\noalign{}
\endhead
\bottomrule\noalign{}
\endlastfoot
Core model families & Qwen 3.5 4B; Gemma 4 E4B; Ministral 3 8B; Granite
4 3B \\
Prespecified larger variants & Qwen 3.5 9B; Gemma 4 12B; Ministral 3
14B \\
Topics & AI-content labels; identity verification for high reach;
chronological-feed defaults; personalized political-ad targeting \\
Seeds & 101, 211, 307, 419 \\
Conditions & topic-only control; peer-ranked feed; single source;
distributed sources \\
Honest agents / rounds & 12 / 6 \\
\end{longtable}

The four-family core contains 256 trials: 4 models x 4 topics x 4 seeds
x 4 conditions. The separately analyzed size extension contains 192
trials: 3 variants x 4 topics x 4 seeds x 4 conditions. Attack direction
is counterbalanced: seeds 101 and 307 push support; seeds 211 and 419
push opposition.

\subsection{Conditions and estimands}\label{conditions-and-estimands}

\textbf{Topic-only control.} Honest agents see the topic and persona but
no peer posts.

\textbf{Peer-ranked feed.} After round 0, each honest agent sees five
previous-round honest posts ranked by peer-generated likes. This
treatment bundles exposure to peer posts with like ranking. Its contrast
with control estimates the effect of the complete feed surface, not the
isolated effect of ranking.

\textbf{Single source.} One adversarial account generates one post per
round from a fixed directional argument prompt. After round 0, every
honest agent sees exactly one adversarial post and four organic posts.

\textbf{Distributed sources.} Four adversarial accounts independently
realize the same directional argument prompt. Their posts are rotated
across viewers. Every honest agent again sees exactly one adversarial
post and four organic posts after round 0.

Thus both attack conditions deliver 60 adversarial impressions per
trial: 12 honest viewers x 5 post-initial rounds. Feed depth, honest
voters, attack direction, argument prompt, and total adversarial
impressions are fixed. The contrast changes source multiplicity and
cross-viewer message realization together; it tests a distributed-source
package rather than the isolated effect of account count.

\subsection{Outcomes}\label{outcomes}

Stance is scored from -3 (strongly oppose) to +3 (strongly support). The
preregistered primary outcome is change in honest-agent alignment with
the attack direction, where $s_{\mathrm{push}} \in \{-1,+1\}$ encodes the
attack direction and $y$ is the agent's discrete stance:

\[
\Delta_{\mathrm{align}} =
\operatorname{mean}(s_{\mathrm{push}} y_{\mathrm{final}})
- \operatorname{mean}(s_{\mathrm{push}} y_{\mathrm{initial}}).
\]

The primary contrast is paired
\texttt{distributed\_sources\ -\ single\_source} within each
model-topic-seed block. The prespecified population-driven influence (PDI)
criterion requires a positive overall contrast, positive estimates in at
least three of four core families, and positive estimates in at least
two thirds of completed topics.

Secondary outcomes include pushed-side share, movement toward the pushed
side, survival of initially opposite-side agents, majority capture, and
final-post lexical similarity. Confirmatory similarity is the mean
off-diagonal cosine similarity among TF-IDF vectors for the 12 honest
agents' final-round posts; vocabulary is fitted within each trial
endpoint.

\subsection{Statistical analysis and
audit}\label{statistical-analysis-and-audit}

The inferential unit is the complete paired model-topic-seed block,
never an individual post. For each contrast we report the paired mean
difference, 20,000-draw percentile block-bootstrap interval, exact
two-sided sign test, and paired sign-flip randomization p-value.
Randomization is exhaustive for at most 20 nonzero pairs and uses one
million deterministic Monte Carlo draws otherwise. Model and topic
strata are released for every outcome.

The PDI stance contrast is the primary test. Feed outcomes are separate
secondary estimands; p-values are nominal and not family-wise adjusted.
We emphasize lexical similarity because its interval excludes zero in
both separately analyzed panels and its direction is positive across all
topic strata.

All 448 trials completed. The artifact contains all 112 four-condition
blocks, 35,616 confirmatory posts, 32,256 honest-agent voter calls, and
448 vote logs. Every exposure audit passes. Post parse failures are
53/20,352 (0.26\%) in the core and 30/15,264 (0.20\%) in the size
extension. Vote parse failures are 140/18,432 (0.76\%) and 145/13,824
(1.05\%), respectively. All rates remain below the frozen 5\% warning
threshold, and failed generations retain prior stance rather than being
silently excluded.

\section{Confirmatory Results}\label{confirmatory-results}

\subsection{Pooled contrasts}\label{pooled-contrasts}

\begin{longtable}[]{@{}
  >{\raggedright\arraybackslash}p{(\linewidth - 8\tabcolsep) * \real{0.1579}}
  >{\raggedleft\arraybackslash}p{(\linewidth - 8\tabcolsep) * \real{0.2105}}
  >{\raggedleft\arraybackslash}p{(\linewidth - 8\tabcolsep) * \real{0.2105}}
  >{\raggedleft\arraybackslash}p{(\linewidth - 8\tabcolsep) * \real{0.2105}}
  >{\raggedleft\arraybackslash}p{(\linewidth - 8\tabcolsep) * \real{0.2105}}@{}}
\toprule\noalign{}
\begin{minipage}[b]{\linewidth}\raggedright
Panel and contrast
\end{minipage} & \begin{minipage}[b]{\linewidth}\raggedleft
n blocks
\end{minipage} & \begin{minipage}[b]{\linewidth}\raggedleft
paired mean
\end{minipage} & \begin{minipage}[b]{\linewidth}\raggedleft
95\% bootstrap CI
\end{minipage} & \begin{minipage}[b]{\linewidth}\raggedleft
randomization p
\end{minipage} \\
\midrule\noalign{}
\endhead
\bottomrule\noalign{}
\endlastfoot
Core: PDI alignment & 64 & +0.0573 & {[}-0.0091, +0.1250{]} & 0.1124 \\
Larger: PDI alignment & 48 & -0.0399 & {[}-0.1128, +0.0347{]} &
0.3317 \\
Core: feed survival & 64 & -0.0391 & {[}-0.0677, -0.0156{]} & 0.0068 \\
Larger: feed survival & 48 & -0.0104 & {[}-0.0312, +0.0035{]} &
0.5000 \\
Core: feed TF-IDF & 64 & +0.0082 & {[}+0.0043, +0.0121{]} & 0.000105 \\
Larger: feed TF-IDF & 48 & +0.0109 & {[}+0.0069, +0.0151{]} &
0.000001 \\
\end{longtable}

The peer-ranked feed produces the only result that replicates clearly
across both panels: final-round honest posts become more lexically
similar. The effect is modest, approximately 0.008 to 0.011 TF-IDF
cosine units, and does not imply that agents reach the same opinion.

Opposite-side survival decreases by 3.9 pp in the core panel, but the
larger variants are inconclusive. The expanded evidence therefore does
not support a model-general minority-suppression law.

The matched-exposure coordination result fails the PDI criterion. The
core estimate is positive but uncertain; the size-extension estimate is
negative and uncertain. Under the implemented treatment, distributing an
equally exposed argument across four sources provides no reliable
general advantage over one source.

\begin{figure}[H]
\centering
\pandocbounded{\includegraphics[keepaspectratio,alt={Pooled preregistered matched-exposure effects}]{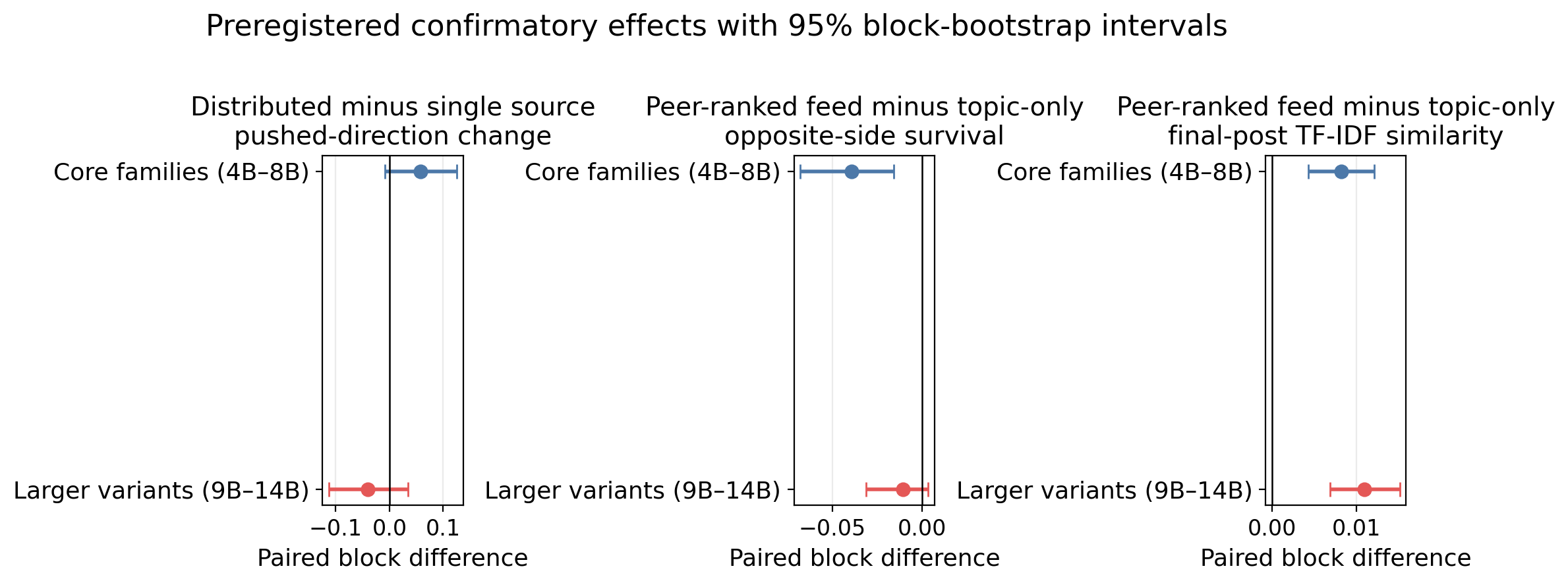}}
\caption{Pooled preregistered matched-exposure effects. Points are paired model-topic-seed block means and horizontal bars are 95\% block-bootstrap intervals; the black vertical line marks no difference. Core families and larger variants are analyzed separately.}
\end{figure}

\subsection{Model and topic
heterogeneity}\label{model-and-topic-heterogeneity}

Lexical-similarity effects are positive for three of four core models
and all three larger variants. Gemma 4 E4B is slightly negative, while
Qwen 3.5 4B and 9B show the largest positive estimates. All four topic
means are positive in each panel. This consistency supports a
feed-convergence result while also showing meaningful model variation in
magnitude.

The core survival effect is concentrated in Qwen 3.5 4B (-12.5 pp).
Gemma 4 E4B is exactly null, and Granite 4 3B and Ministral 3 8B have
smaller estimates. In the larger panel, Qwen 3.5 9B is negative while
Gemma 4 12B and Ministral 3 14B are near zero or positive. The
appropriate interpretation is model dependence, not a universal effect.

For PDI, three of four core-family means are positive, but only two of
four topic means are positive and the overall interval crosses zero. In
the size extension, only one of three model means and one of four topic
means are positive. The larger Qwen variant is significantly negative
within its 16-block stratum, illustrating why a pooled positive
exploratory result cannot be generalized across model variants.

\begin{figure}[H]
\centering
\pandocbounded{\includegraphics[keepaspectratio,alt={Model-stratified preregistered matched-exposure effects}]{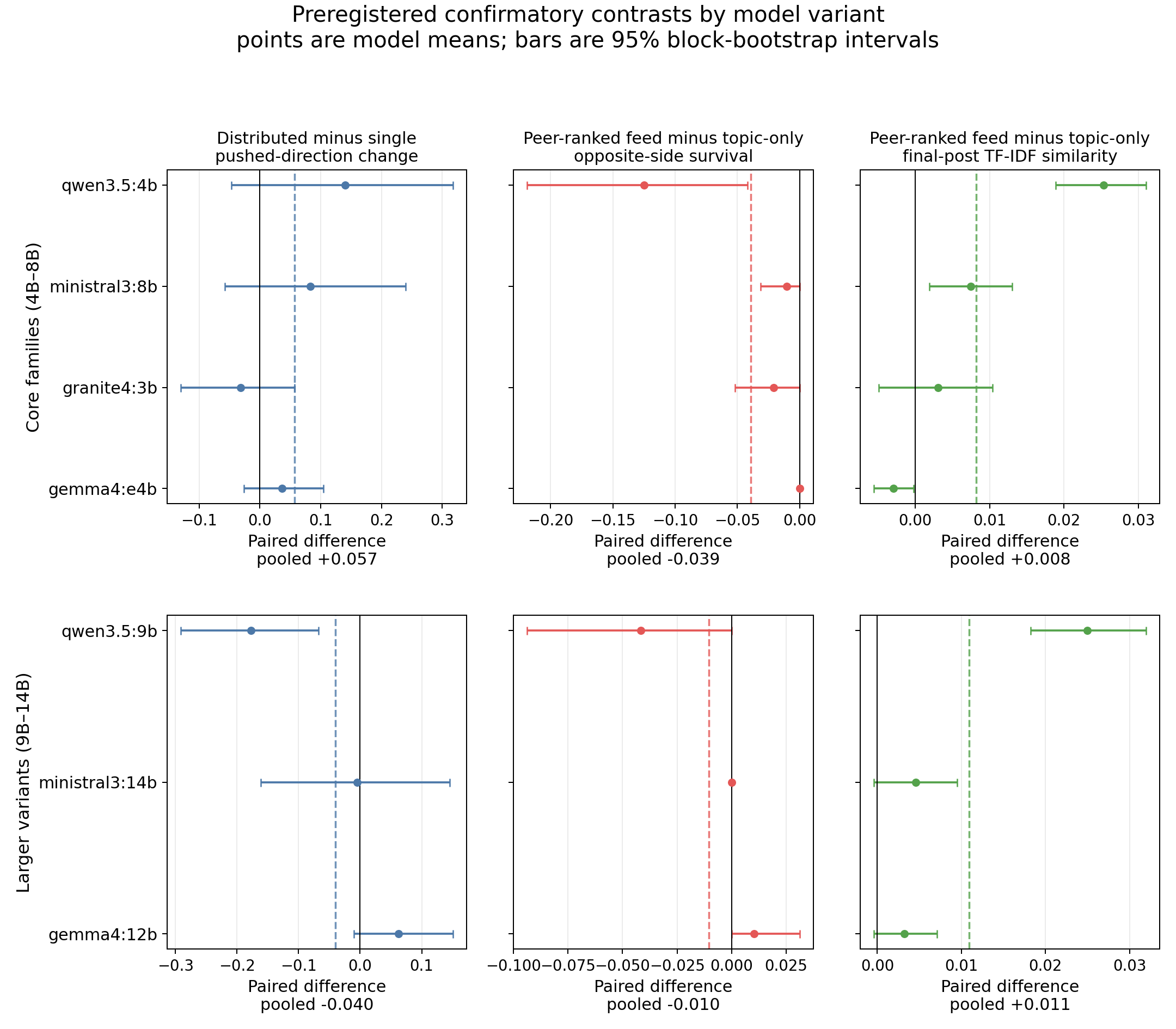}}
\caption{Model-stratified preregistered matched-exposure effects. Each point is a model-specific paired mean and each horizontal bar is a 95\% block-bootstrap interval. The solid black line marks no difference; the dashed colored line marks the pooled estimate for that panel and outcome.}
\end{figure}
\FloatBarrier

\section{Exploratory Origin and Measurement
Audit}\label{exploratory-origin-and-measurement-audit}

Before the confirmation, an 80-trial exploratory stage used Qwen 3.5 2B
and Llama 3.2 3B on one platform-policy topic. It screened ordinary
feedback surfaces, coordinated-account load, one-account rank
amplification, and misinformation interventions. Those experiments
generated the two confirmatory questions: whether peer-feedback
convergence survives broader replication and whether distributed sources
outperform one source after exposure is matched. Their small cells (two
models x two seeds) and unmatched attack mechanisms are not used as
confirmatory evidence.

The exploratory misinformation probe yielded a separate measurement
warning. An all-keyword detector recorded zero honest-agent matches
despite close restatements in the raw posts, a failure we call
\textbf{paraphrase leakage}. Polarity-blind embedding similarity then
conflated endorsement with rebuttal, which we call \textbf{stance
confounding}. Two independent LLM stance judges agreed on only 43.6\% of
endpoint labels (Cohen's kappa=0.258) and disagreed on treatment
direction. Consequently, the paper does not rank misinformation
interventions. These observations motivate human annotation or a
validated stance-aware outcome; they do not establish that production
moderation systems generally fail.

All exploratory traces remain in the release because they document
hypothesis formation and unsuccessful measurement approaches. Keeping
them separate from the frozen confirmation prevents the larger dataset
from retroactively turning exploratory choices into preregistered
claims.

\section{Discussion}\label{discussion}

\subsection{What the results
establish}\label{what-the-results-establish}

The strongest result is narrow and reproducible: under this testbed,
exposure to a feed of prior peer posts ranked by endogenous likes
increases lexical similarity relative to seeing only the topic. In agent
populations, the feed is not merely a display surface; it is part of the
system being evaluated. The next mechanism-isolation study should add an
unranked peer-feed arm to separate exposure from ranking.

The coordination result is equally useful because it is negative. When
adversarial impressions are fixed, the tested four-source package does
not reliably move stance more than one source. This does not imply that
coordination is harmless. Real campaigns can gain reach, vary arguments,
target subgroups, occupy network bottlenecks, and exploit human social
identity. It shows that these mechanisms must be modeled explicitly
rather than attributed to account count alone.

\subsection{Implications for
evaluation}\label{implications-for-evaluation}

Three design principles follow:

\begin{enumerate}
\def\labelenumi{\arabic{enumi}.}
\tightlist
\item
  \textbf{Match exposure before comparing threat mechanisms.} Otherwise
  source count, reach, and message volume are inseparable.
\item
  \textbf{Report model strata.} Aggregate LLM-population effects can be
  driven by one family or reverse with model size.
\item
  \textbf{Validate semantic outcomes.} Surface matching, embeddings, and
  unvalidated LLM judges answer different questions and should not be
  substituted after outcomes are observed.
\end{enumerate}

\subsection{Limitations}\label{limitations}

\textbf{Synthetic populations.} PV-SST does not model human emotion,
account deletion, long-term relationships, off-platform information, or
platform-specific recommendation systems. No magnitude is a prediction
for people, Snapchat, Instagram, X, or another named platform.

\textbf{Bundled feed treatment.} The topic-only control and peer-ranked
feed differ in both peer-post exposure and ranking. The experiment
establishes a feed effect, not a ranking-only effect.

\textbf{Bundled distributed treatment.} Four sources generate
independently realized versions of one argument prompt, whereas the
single source generates one realization per round. The contrast tests
this distributed package under equal impressions, not account count in
isolation.

\textbf{Model and topic scope.} The confirmation covers four open-weight
families and three larger variants, not proprietary frontier models. All
four topics concern platform policy; health, elections, identity,
entertainment, and less deliberative content may differ.

\textbf{Crossed simulation blocks.} Bootstrap and randomization tests
treat paired model-topic-seed blocks as exchangeable simulation
replications. They are not 112 independent platforms. Model and topic
strata expose heterogeneity but do not establish population-level
inference beyond the tested matrix.

\textbf{Secondary-outcome multiplicity.} Feed outcomes are prespecified
secondary estimands with nominal p-values. The lexical result is
emphasized because it replicates in both panels and across topic strata,
not because one favorable cell was selected.

\textbf{Human validation.} The released one-shot majority-cue diagnostic
shows large model heterogeneity but has no matched human treatment. A
planned ChangeMyView replay was implemented only with synthetic
placeholders. Simulation-to-human transfer remains uncalibrated.

\section{Conclusion}\label{conclusion}

A seven-variant, four-topic, four-seed confirmation provides a
falsifiable evaluation of this LLM-agent population. A peer-ranked feed
increases final-post lexical similarity
across both model-size panels. Minority-view suppression is not
model-general. Distributed sources gain no reliable stance-moving
advantage over one source when adversarial impressions are held fixed.

The broader lesson is methodological. LLM-agent platform risks should be
decomposed into feed exposure, ranking, source multiplicity, message
diversity, and reach, then tested with paired run-level designs.
Generated posts are evidence to audit, not independent samples to
inflate statistical power. Human-platform claims require human evidence.

\section*{Acknowledgments}

This work was conducted independently. Experiment orchestration, statistical
auditing, figure rendering, and manuscript drafting used AI-assistant support
under direct human supervision. The authors are responsible for the research
design, claim verification, and final manuscript. Cloud compute was rented
from RunPod; the confirmatory matrix ran on one NVIDIA RTX 5090.

\IfFileExists{main.bbl}{
}{\bibliographystyle{plainnat}\bibliography{references}}

\end{document}